\documentclass[fp,twocolumn]{jpsj3}
\usepackage{txfonts}
\usepackage{mathrsfs}
\usepackage{color}
\newcommand{\lsim}
 {\ \raise.35ex\hbox{$<$}\kern-0.75em\lower.5ex\hbox{$\sim$}\ }
\newcommand{\gsim}
 {\ \raise.35ex\hbox{$>$}\kern-0.75em\lower.5ex\hbox{$\sim$}\ }
\def\journal #1#2#3#4{#1 {\bf #2}, #3 (#4)}
\def\APNY{Ann.\ Phys.\ (New York)}

\def\JPConfS{J.\ Phys.:\ Conf.\ Ser.}

\def\JPSJ{J.\ Phys.\ Soc.\ Jpn.}

\def\NC{Nat.~Commun.}
\def\NM{Nat.~Mater.}

\def\PRB{Phys.\ Rev.\ B}

\def\PRL{Phys.\ Rev.\ Lett.}

\def\RMP{Rev.\ Mod.\ Phys.}
\def\SA{Sci.~Adv.}
\def\SST{Supercond.~Sci.~Technol.}

\title{Excited States beyond Mott Gap in Half-Filled-Band Hubbard Model} 

\author{
Hisatoshi Yokoyama$^1$\thanks{yoko@cmpt.phys.tohoku.ac.jp}, 
Kenji Kobayashi$^2$, Tsutomu Watanabe$^2$, and 
Masao Ogata$^3$}
\inst{
$^1$Department of Physics, Tohoku University, Aoba-ku, Sendai, 980-8578, 
Japan \\
$^2$Department of Natural Science, Chiba Institute of Technology, Narashino, 
275-0023, Japan \\
$^3$Department of Physics and Trans-scale Quantum Science Institute, 
University of Tokyo, Bunkyo, Tokyo 113-0033, Japan
} %\\

\abst{
In connection with recent experiments on excitation in which Mott 
insulators change to conductors, we study the properties of excited states 
beyond the Mott gap as quasi-stationary states for a two-dimensional 
Hubbard ($t$-$t'$-$U$) model at half filling. 
A variational Monte Carlo method is used with trial wave functions for  
paramagnetic or normal (PM), superconducting with $d_{x^2-y^2}$-wave ($d$-SC), 
isotropic $s$-wave, and extended $s$-wave symmetries, and 
antiferromagnetic (AF) states. 
The excited states are generated by imposing a minimum number of doubly 
occupied sites (doublons) $D_{\rm L}$ on the lowest-energy states. 
For $U\gtrsim W$ ($W$: band width), $d_{\rm L}=D_{\rm L}/N_{\rm s}$ 
($N_{\rm s}$: number of sites) corresponds to the excitation intensity. 
It is found that the AF state is the most stable among the states we treated 
for $d_{\rm L}\lesssim 0.14$ and insulating. 
The PM and $d$-SC states become conductive over a threshold $d_{\rm Lc}$, 
and the conduction is caused by unbound doublons and holons (empty sites). 
The PM state arises for $d_{\rm L}\gtrsim 0.14$, but the $d$-SC state is 
always hidden by the AF state. 
The $s$-wave-type superconducting states are not stabilized for any 
parameter set.  
}

\begin{document}
\maketitle

\section{Introduction\label{sec:intro}}
Recently, Mott-insulator-to-conductor transitions induced by ultrafast 
photoexcitation \cite{Iwai,Okamoto,Miyamoto,Terashige} and pulse electric 
fields\cite{EF} with $\Delta E\gtrsim U$ ($\Delta E$: excitation energy, 
$U$: onsite Coulomb repulsion or Mott gap) have been intensively studied with 
application to high-speed switching devices in mind. 
A similar experiment was also performed in cold-atom systems.\cite{Strohmaier} 
To date, related theoretical studies had shed light mostly on the dynamical 
aspects after such 
excitation.\cite{Aoki,Prelovsek,Iyoda,Yonemitsu,Shinjo,Werner,eta} 
The state after excitation develops or relaxes with ultra-high speeds, 
but it is also significant from a basic point of view to treat the states 
as quasi-stationary states, as previously studied using $t$-$J$-type models
\cite{Takahashi1,Takahashi2,Gomi}. 
It is still nontrivial whether the ``fresh" excited state (shortly after 
pumping) remains insulating or has changed to conductive. 
\par

In this study, we assume that the states excited beyond the Mott gap, in 
which extra doublons (doubly occupied sites, D) and holons (empty sites, H) 
are generated, are stationary for sufficiently long time for electronic 
processes. 
This assumption is not necessarily irrelevant, because relaxation processes 
through phonons and light emission can have much longer time scales.
Here, we study these excited states as steady states using a two-dimensional 
Hubbard model at half filling by means of a variational Monte Carlo (VMC) 
method,\cite{SY} and reveal whether the stable excited state to be realized 
is insulating or conductive and what type of order prevails in it. 
In the VMC processes, small-scale energy ($E\ll U$) dissipates to 
the outside of the system, and it reaches the optimized state within 
the given condition of D mentioned below, in contrast with 
energy-conservative schemes. 
\par

As research in this line, various aspects of the states with fixed small 
numbers of doublons ($\mathscr{D}=1$ -- $4$) were studied by applying an 
exact-diagonalization method to small clusters ($\sim 20$ sites) of 
a $t$-$J$-type effective Hamiltonian without\cite{Takahashi1,Takahashi2} 
and with\cite{Gomi} nearest-neighbor repulsive interaction. 
These pioneering studies showed important aspects of excited states, 
for instance, an antiferromagnetic (AF) order survives for weak excitation
intensity, and repulsive correlation works between nearest-neighbor D-H 
pairs, as we will refer to. 
However, important problems have been left untouched. 
For instance, system-size dependence, which is crucial especially for 
$d_{x^2-y^2}$-wave superconductivity,\cite{YOTKT} is not easy to be checked. 
It is not clear how the effective Hamiltonian reproduces the properties of 
the original Hubbard model, for example, the effects of D and H already 
existing in the ground state were not clarified on the excited states. 
The properties of higher-energy states with $\Delta E\gtrsim U$, such as 
a $d_{x^2-y^2}$-wave superconducting ($d$-SC) state, are still unknown. 
The present VMC scheme can shed light on these points. 
\par

To represent the excited states, we intentionally introduce additional 
doublons and holons into the trial states by prohibiting the total number 
of doublons (${\cal D}$) from being smaller than the lower bound 
$D_{\rm L}$ we set (${\cal D}\ge D_{\rm L}$). 
This operation is exactly and easily carried out using the VMC method, 
and was found virtually equal to the creation of additional $D_{\rm L}$ 
doublons for $U\gtrsim U_{\rm c}\sim W$ ($U_{\rm c}/t$: the Mott transition 
point, $W$: band width). 
Thus, $D_{\rm L}$ and $D_{\rm L}/N_{\rm s}\equiv d_{\rm L}$  ($N_{\rm S}$: 
number of sites) corresponds to $\mathscr{D}$ in the above studies 
for the $t$-$J$-type model and to the average intensity of photon absorption 
in the above photoexcitation experiments,\cite{Okamoto,Miyamoto,Terashige} 
respectively. 
We apply this operation to a paramagnetic (PM) or normal state, an 
AF state, and superconducting (SC) states of three kinds 
of pairing symmetry, a $d_{x^2-y^2}$ wave, an isotropic $s$ wave ($s$-SC) 
and an extended-$s$ wave (x-SC), and study the properties of each state 
and mutual stability, in particular as functions of 
$d_{\rm L}$ [or equivalently of doublon density $d$ in Eq.~(\ref{eq:D})]. 
\par

It is intriguing to know whether the properties of the lowest-energy 
states\cite{YOTKT,SY}  are preserved or change to different features 
in the excited states. 
For example, (i) all of the above states for $U>U_{\rm c}$ or 
$U>U_{\rm AF}$ ($U_{\rm AF}/t$: AF transition point) are insulating 
in the lowest-energy cases. 
Are overabundant doublons and holons make the states conductive? 
(ii) the lowest-energy $d$-SC state exhibits robust pairing magnitude 
$P_d$ immediately below $U_{\rm c}/t$ ($\sim 6.6$) and loses it in the Mott 
insulating regime ($U>U_{\rm c}$). 
Does the excitation enhance (revive) or lower $P_d$ ($T_{\rm c}$)? 
(iii) Is the $d$-SC state, which is the most stable among the lowest-energy 
SC states, defeated by $s$-wave-type states, to which the $\eta$-pairing 
state\cite{eta} belongs, for $D_{\rm L}>0$? 
\par

This article is organized as follows: 
In Sect.~\ref{sec:formalism}, the model and method we use are introduced
(Sect.~\ref{sec:method}), and the relation $D_{\rm L}\sim\mathscr{D}$ for 
large $U/t$ is confirmed (Sect.~\ref{sec:DL}).
In Sect.~\ref{sec:PM}, we consider the excited states in the PM branch. 
In Sects.~\ref{sec:SC} and \ref{sec:AF}, the SC states and AF state are 
studied, respectively. 
In Sect.~\ref{sec:summary}, we recapitulate the main results. 
In Appendix, we discuss the effects of diagonal hopping term.
Preliminary results for $d$-SC state have been published in a 
proceedings.\cite{proc}
\par

%==========================================
\section{Formalism\label{sec:formalism}}
%==========================================
\subsection{Model and Method\label{sec:method}}
For addressing excited states with $\Delta E>U$, the single-band 
Hubbard model ($U\ge 0$) is suitable. 
We consider the case on a square lattice with diagonal transfer: 
\begin{align}
{\mathscr H}&={\mathscr H}_{\rm kin}+{\mathscr H}_U 
\nonumber \\
&=-\sum_{(i,j),\sigma}t_{ij}
             \left(c^\dagger_{i\sigma}c_{j\sigma} + \mbox{H.c.}\right) 
             +U\sum_j n_{j\uparrow}n_{j\downarrow},
\label{eq:Hamil}
\end{align} 
where %$d_j=n_{j\uparrow}n_{j\downarrow}$, 
$n_{j\sigma}=c^\dag_{j\sigma}c_{j\sigma}$ and $(i,j)$ indicates the pairs 
on sites $i$ and $j$. 
We set the hopping integral $t_{ij}$ as $t$ ($\ge 0$) for nearest neighbors, 
$t'$ for diagonal (next-nearest) neighbors, and $0$ otherwise 
(${\mathscr H}_{\rm kin}={\mathscr H}_t+{\mathscr H}_{t'}$). 
As discussed in Appendix, $t'/t$ dependence is undetectable in the AF state 
and not essential for the $d$-SC and PM states at least for 
$|t'/t|\lesssim 0.5$ and relevant excitation strength 
($D_{\rm L}/N_{\rm s}\lesssim 0.08$). 
Hence, we fix $t'/t$ at $-0.3$ (a typical value for cuprate superconductors) 
in the main text. 
We focus on the half-filled band ($n=N/N_{\rm s}=1$ or $\delta=|1-n|=0$, 
$N$: number of electrons). 
We use $t$ and the lattice spacing as the units of energy and length, 
respectively. 
\par

To this model, we apply a variational Monte Carlo (VMC) method, 
\cite{YS1,Umrigar,SY} which enables us to exactly treat many-body wave 
functions and continuously connects weakly and strongly correlated 
regimes even if some phase transition lies between them. 
To construct trial excited states in which extra doublons and holons are 
induced, we extend the Jastrow form previously used for the lowest-energy 
state ($D_{\rm L}=0$) $\Psi_0$\cite{SY} to 
\begin{equation}
\Psi_{D_{\rm L}}~(=\Psi_{d_{\rm L}})={\cal P}_{D_{\rm L}}\Psi_0, 
\label{eq:excite}
\end{equation}
where ${\cal P}_{D_{\rm L}}$ is a projector that imposes the condition 
${\cal D}\ge D_{\rm L}$. 
\par

Before explaining ${\cal P}_{D_{\rm L}}$, we review $\Psi_0$ 
($={\cal P}\Phi={\cal P}_{\rm G}{\cal P}_Q\Phi$). 
${\cal P}_{\rm G}$ is the well-known Gutzwiller (onsite) 
projector:\cite{Gutz} 
\begin{equation}
{\cal P}_{\rm G}=\prod_j\left[1-(1-g)n_{j\uparrow}n_{j\downarrow}\right], 
\end{equation}
with a parameter $g$, ${\cal P}_Q$ is a nearest-neighbor D-H binding 
factor\cite{YS3,YOTKT} crucial for Mott physics: 
${\cal P}_Q=\prod_j\left(1-Q_j\right)$, where 
\begin{equation}
Q_j=\zeta_{\rm d}~d_j\prod_\tau(1-h_{j+\tau})
         +\zeta_{\rm h}~h_j\prod_\tau(1-d_{j+\tau}), 
\label{eq:SymQ} 
\end{equation}
$d_j=n_{j\uparrow}n_{j\downarrow}$, 
$h_j=(1-n_{j\uparrow})(1-n_{j\downarrow})$, $\zeta_{\rm d}$ and 
$\zeta_{\rm h}$ are D-H binding parameters, and $\tau$ runs over all the 
nearest-neighbor sites of site $j$. 
At half filling, a relation $\zeta_{\rm d}=\zeta_{\rm h}$ ($\equiv\zeta$) 
holds owing to the electron-hole symmetry. 
\par

We turn to the one-body (determinantal) part 
$\Phi$.\cite{mixed}
As a PM or normal state $\Phi_{\rm PM}$, we employ a Fermi sea 
\begin{equation}
\Phi_{\rm PM}=\prod_{{\bf k}~\in~\{{\bf k}\}_{\rm occ},~\sigma}
c^\dag_{{\bf k}\sigma}|0\rangle, 
\label{eq:FS}
\end{equation}
where ${\bf k}$ is inside the renormalized Fermi surface 
$\{{\bf k}\}_{\rm occ}$; as explained below, four band parameters 
$t_1$--$t_4$ are implicitly used to determine $\{{\bf k}\}_{\rm occ}$. 
For the SC state, we use an $N$-electron BCS wave function with typical 
pairing-gap symmetries $\lambda$ ($=d$, $s$, or x),  
\begin{align}
\Phi_\lambda &=\left(\sum_{\bf k}\phi_{\bf k}
c_{{\bf k}\uparrow}^\dagger c_{{\bf -k}\downarrow}^\dagger
\right)^\frac{N}{2}|0\rangle, 
\label{eq:Phi_d} \\
&\phi_{\bf k}=\frac{v_{\bf k}}{u_{\bf k}}=\frac{\Delta_{\bf k}}
{\varepsilon_{\bf k}-\mu+
\sqrt{(\varepsilon_{\bf k}-\mu)^2+\Delta_{\bf k}^2}}, 
\label{eq:BCSDelta}
\end{align}
\begin{equation}
\Delta_{\bf k}=\left\{
\begin{array}{ll}
\Delta_d(\cos k_x-\cos k_y) & \quad d_{x^2-y^2}\mbox{-wave} \\
\Delta_s & \quad \mbox{isotropic}\ s\mbox{-wave} \\
\Delta_{\rm x}(\cos k_x+\cos k_y) & \quad \mbox{extended}\ s\mbox{-wave}
\label{eq:symmetry}
\end{array}
\right., 
\end{equation}
where $\Delta_d$, $\Delta_s$, and $\Delta_{\rm x}$ represent pairing 
magnitude (not necessarily indicating coherence strength\cite{YOTKT}) of 
the respective symmetries, and are to be optimized. 
$\mu$ is a parameter, which is reduced to the chemical potential for 
$U/t\rightarrow 0$. 
For the AF state, a simple Hartree-Fock solution is used, 
\begin{align}
\Phi_{\rm AF}&=\prod_{\{{\bf k}\}_{\rm occ},~\sigma}
a_{{\bf k}\sigma}^\dag\left|0\right>, \\
&a^\dag_{{\bf k},\sigma}=\alpha_{\bf k} c^\dag_{{\bf k},\sigma}+
\mbox{sgn}(\sigma)\ \beta_{\bf k} c^\dag_{{\bf k}+{\bf Q},\sigma}, 
\label{eq:QP1} \\
&a^\dag_{{\bf k}+{\bf Q},\sigma}=-\mbox{sgn}(\sigma)\ \beta_{\bf k} 
c^\dag_{{\bf k},\sigma}+\alpha_{\bf k} c^\dag_{{\bf k}+{\bf Q},\sigma}, 
\label{eq:QP2} 
\end{align}
where ${\bf Q}$ is the AF nesting vector $(\pi,\pi)$, 
$\mbox{sgn}(\sigma)=1$ ($-1$) for $\sigma=\uparrow$ ($\downarrow$), and
\begin{equation}
\alpha_{\bf k}\ (\beta_{\bf k})=\frac{1}{\sqrt{2}}
\sqrt{1-(+)\frac{\varepsilon_{\bf k}} 
       {\left(\varepsilon_{\bf k}\right)^2+\Delta_{\bf AF}^2}}. 
\label{eq:alpha-beta} 
\end{equation}
$\Delta_{\rm AF}$ corresponds to an AF gap parameter in the sense of 
the mean-field theory, but is renormalized owing to ${\cal P}$ here. 
\par

In each $\Phi$, a band renormalization effect (BRE) is introduced 
by optimizing the tight-binding band $\varepsilon_{\bf k}$, which is 
expanded up to sixth-neighbor sites \cite{SY,note-fifth}: 
\begin{align}
\varepsilon_{\bf k}=\gamma_{\bf k}&
  +\varepsilon_1({\bf k})+\varepsilon_2({\bf k}) 
  +\varepsilon_3({\bf k})+\varepsilon_4({\bf k}), 
\label{eq:BR} \\
\gamma_{\bf k}&=-2t(\cos k_x+\cos k_y), \tag{\ref{eq:BR}a} \\
\varepsilon_1({\bf k})&=-4t_1\cos k_x\cos k_y, \tag{\ref{eq:BR}b} \\
\varepsilon_2({\bf k})&=-2t_2(\cos 2k_x+\cos 2k_y), \tag{\ref{eq:BR}c} \\
\varepsilon_3({\bf k})&=
    -4t_3(\cos 2k_x\cos k_y+\cos k_x\cos 2k_y), \tag{\ref{eq:BR}d} \\
\varepsilon_4({\bf k})&=-2t_4(\cos 3k_x+\cos 3k_y). \tag{\ref{eq:BR}e}
\end{align}
Such BRE is crucial for considering PM and, especially, AF 
states.\cite{YOTKT,SY}
The optimized band parameters $t_1$--$t_4$ generally become different 
among $\Phi$. 
Because the parameters in $\Phi$ ($t_1$--$t_4$, $\mu$, $\Delta_d$) have 
considerable redundancy, they are not optimized at unique values; however, 
the correlation parameters ($g$ and $\zeta$) as well as the minimal 
energy and corresponding physical quantities are uniquely determined 
within statistical errors. 
\par

\begin{table}
\caption{
Doublon density $d$ in lowest-energy states ($\Psi_0$, $d_{\rm L}=0$) 
for relevant three phases in half-filled-band square lattice with 
$|t'/t|\le 0.5$. 
For information, the data of a staggered flux (SF) phase\cite{SF} are 
added. 
The data with * indicate that the optimized state is (super)conducting; 
the Mott\cite{YOT} (AF, PM-SF) transition point $U_{\rm c}/t$ 
($^\dag U_{\rm AF}/t$, $^\# U_{\rm SF}/t$) is shown in the last column.
The values for $N_{\rm s}=14\times 14$ are shown. 
}
\begin{center}
\begin{tabular}{c|rrrrr|r}
\hline
$\Psi$ &       &     & $d_0$~~~ &      &      & $U_{\rm c}/t$ 
\\
       & $U/t=~~6$ & $8$ & $10$ & $12$ & $20$ & 
\\
\hline
PM     & $^* 0.122$ & $^*0.055$ & $0.019$ & $0.015$ & $0.007$ & $\sim 8.5$
\\
$d$-SC & $^*0.123$ & $0.032$ & $0.026$ & $0.020$ & $0.010$ & $6.6$ 
\\
AF     & $0.083$ & $0.055$ & $0.038$ & $0.027$ & $0.011$ & $^\dag3.0$ 
\\
\hline
SF     & $0.109$ & $0.031$ & $0.023$ & $0.017$ & $0.009$& $^\#5.0$
\\
\hline
\end{tabular}
\end{center}
\vskip -4mm
\label{table:BR}
\end{table}

It is known\cite{YTOT,YOT,DMFT,SY} that each $\Psi_0$ is (super)conducting for 
small $U/t$, brings about a first-order conductor-insulator transition 
at $U=U_{\rm c}$ (Mott-transition point) for PM and SC states or 
at $U=U_{\rm AF}$ (AF transition point) for AF state (see Table 
\ref{table:BR}), and becomes insulating for $U>U_{\rm c}$ or $U_{\rm AF}$. 
We take notice of excitation in the insulating regimes below, although we 
will also mention aspects near the transition points. 
\par 

\begin{figure}[t]
\vskip -0.2cm
%\hskip -0.2cm
\begin{center}
\includegraphics[width=8.5cm]{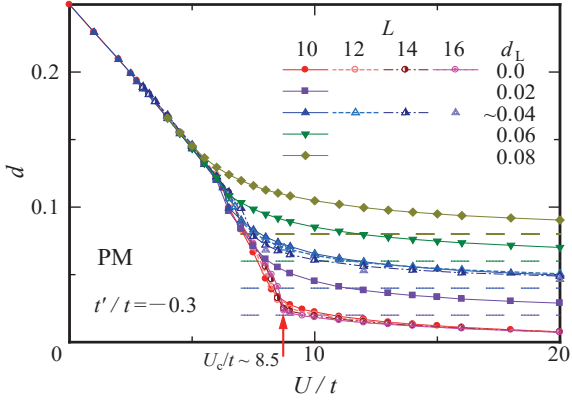}
\end{center}
\vskip -0.4cm
\caption{(Color online) 
$U/t$ dependence of doublon density for some values of $d_{\rm L}$ 
and system size $L$ estimated by VMC using $\Psi_{\rm PM}$. 
Guide lines of $d=d_{\rm L}$ are shown with long-dashed lines with the 
corresponding colors, to each of which $d$ should converge for 
$U/t\rightarrow\infty$. 
The red arrow on the $d_{\rm L}=0$ curve indicates the Mott-transition 
point $U_{\rm c}/t$. 
}
\vskip -0.5cm
\label{fig:d-vs-u-PM}
\end{figure} 
Now, we explain the projector ${\cal P}_{D_{\rm L}}$ in 
Eq.~(\ref{eq:excite}) for D-H excitations. 
It is known that the doublon density, 
\begin{equation}
d=\frac{D}{N_{\rm s}}=\frac{1}{N_{\rm s}}
\sum_j\langle n_{j\uparrow}n_{j\downarrow}\rangle,
\label{eq:D}
\end{equation}
in the lowest-energy state $\Psi_0$ [denoted by ($d=$)~$d_0$], is finite 
even in the insulating regime, 
as shown in Table \ref{table:BR}. 
In this regime, however, the doublons are tightly bound to the counter 
holons; therefore, there is no free charge carrier.\cite{Miyagawa} 
As $U/t$ increases, $d$ decreases as $\propto t/U$ and completely vanishes
in the limit of $U/t\rightarrow\infty$ (red curves in 
Fig.~\ref{fig:d-vs-u-PM}). 
By applying ${\cal P}_{D_{\rm L}}$, we force $\Psi_{D_{\rm L}}$ to always 
have at least $D_{\rm L}$ doublons, where $D_{\rm L}$ is given in each 
calculation; in other words, ${\cal P}_{D_{\rm L}}$ 
restricts the space of $\Psi_0$ to ${\cal D}\ge D_{\rm L}$, where ${\cal D}$ 
indicates the number of doublons in an electron configuration. 
Therefore, $\Psi_{D_{\rm L}=0}$ indicates the original $\Psi_0$, and 
$\Psi_{D_{\rm L}}$ for $U/t=\infty$ is the state in which correctly 
$D_{\rm L}$ doublons (and holons) exist in any configuration: 
${\cal D}=D_{\rm L}$. 
For comparing different system sizes, it is convenient to use the lowest 
doublon density $d_{\rm L}\equiv D_{\rm L}/N_{\rm s}$ instead of the 
lowest number of doublons $D_{\rm L}$. 
Similarly, we often write $\Psi_{d_{\rm L}}$ (${\cal P}_{d_{\rm L}}$) 
for $\Psi_{D_{\rm L}}$ (${\cal P}_{D_{\rm L}}$)  in Eq.~(\ref{eq:excite}). 
\par

\begin{figure}[t]
%\vskip 0.2cm
%\hskip -0.2cm
\begin{center}
\includegraphics[width=8.5cm]{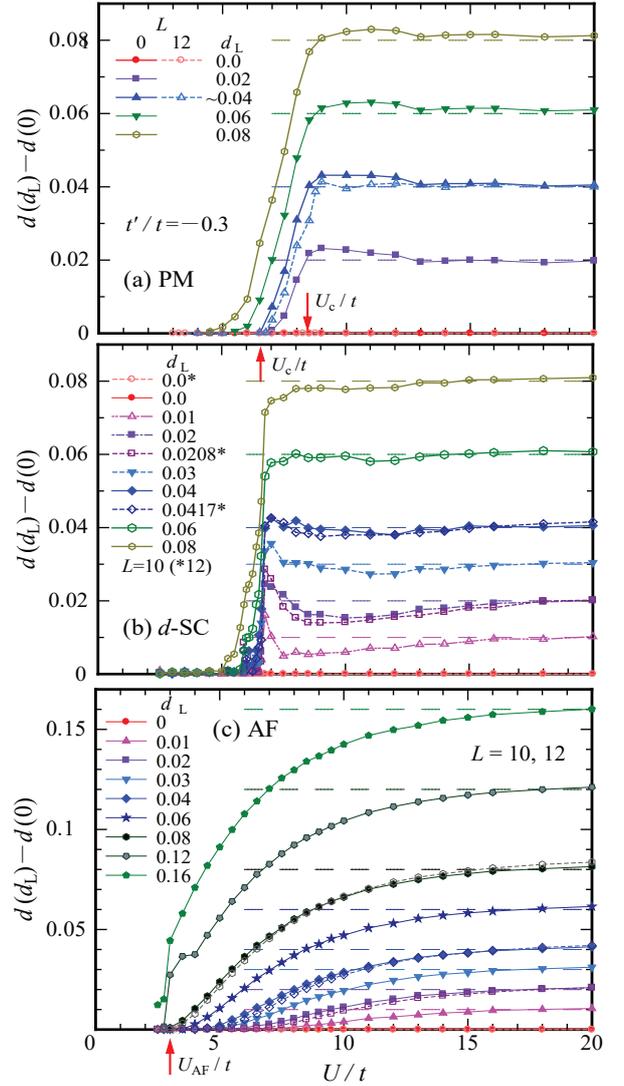}
\end{center}
\vskip -0.2cm
\caption{(Color online) 
Increment of doublon density [$\Delta d=d(d_{\rm L})-d(0)$] for various 
values of $d_{\rm L}$ as functions of $U/t$ for (a) PM, (b) $d$-SC, and 
(c) AF states. 
The long-dashed lines in each panel indicates guide lines satisfying 
$\Delta d=d_{\rm L}$ for respective $d_{\rm L}$. 
In (c), similarly to (b), data for $L=12$ are shown with open symbols and 
dashed lines for $d_{\rm L}\sim 0.02$, $0.04$, and $0.08$. 
The transition point in each state is indicated by a red arrow on a 
horizontal axis. 
}
\label{fig:d-diff}
\end{figure} 

We compute expectation values with respect to $\Psi_{d_{\rm L}}$ using 
a VMC procedure similar to that used in Ref.~\citen{SY}. 
One can easily deal with ${\cal P}_{d_{\rm L}}$ in this procedure. 
We use systems of $N_{\rm s}=L\times L$ sites of $L=10$ -- $16$ with 
the periodic-antiperiodic boundary conditions. 
In a sweep, we iterate processes of optimization typically 160 times for 
each parameter with $5\times10^4$ samples. 
We calculate the variational energy per site, 
\begin{equation}
E=\frac{1}{N_{\rm s}}
\frac{\langle\Psi_{d_{\rm L}}|{\mathscr H}|\Psi_{d_{\rm L}}\rangle}
{\langle\Psi_{d_{\rm L}}|\Psi_{d_{\rm L}}\rangle}, 
\label{eq:E}
\end{equation}
with $2$--$40$ (typically $6$--$10$) different initial values (sweeps), 
because calculations are often trapped in metastable values near the 
global minimum,\cite{note-min} 
and adopt the result with the lowest $E$ and reasonable statistical errors 
as the optimized $\Psi_{d_{\rm L}}$. 
Using this $\Psi_{d_{\rm L}}$, we estimate expectation values of various 
quantities with $5\times10^4$ samples. 
\par

\subsection{$d_{\rm L}$ as increment of doublon density\label{sec:DL}}
In advance, we show by actual calculations that $d_{\rm L}$ approximately 
corresponds to the increment of $d$ in excitations ($\Delta d$). 
Figure \ref{fig:d-vs-u-PM} shows how $U/t$ dependence of doublon density 
evolves as $d_{\rm L}$ is increased for the PM state. 
Let us write the doublon density for $d_{\rm L}$ as $d(d_{\rm L})$. 
In the case of $d\gg d_{\rm L}$ in the metallic regime ($U<U_{\rm c}$), 
$d(d_{\rm L})$ is almost independent of $d_{\rm L}$, 
because $\Psi_0$ seldom includes configurations with ${\cal D}<D_{\rm L}$. 
In the insulating regime ($U>U_{\rm c}$), $d$ seems to increase in 
proportion to $d_{\rm L}$. 
To check this point, we show $\Delta d$ [$\equiv d(d_{\rm L})-d(0)$] 
in Fig.~\ref{fig:d-diff} when various values of $d_{\rm L}$ is imposed 
for the three states. 
For $\Psi_{d_{\rm L}}^{({\rm PM})}$ and $\Psi_{d_{\rm L}}^{(d{\rm -SC})}$, 
the given $d_{\rm L}$ approximately leads 
to the increase in $d$ (except for small $d_{\rm L}$). 
For $\Psi_{d_{\rm L}}^{({\rm AF})}$, $\Delta d(d_{\rm L})$ gradually 
approaches $d_{\rm L}$ 
as $U/t$ increases, probably because the character of $\Psi_{\rm AF}$ 
as an insulator gradually varies from the Slater type to the Mott type.  
Anyway, we can consider that the relation $\Delta d=d_{\rm L}$ is 
approximately realized in the Mott regime  ($U>U_{\rm c}\sim W$, $W$: 
band width) in the three states; as mentioned, $\Delta d$ is proportional 
to the excitation intensity (e.g., optical intensity) in experiments. 
\par

%==========================================
\section{Paramagnetic (Normal) States\label{sec:PM}}
\begin{figure}[t]
%\vskip 0.2cm
%\hskip -0.2cm
\begin{center}
\includegraphics[width=8.5cm]{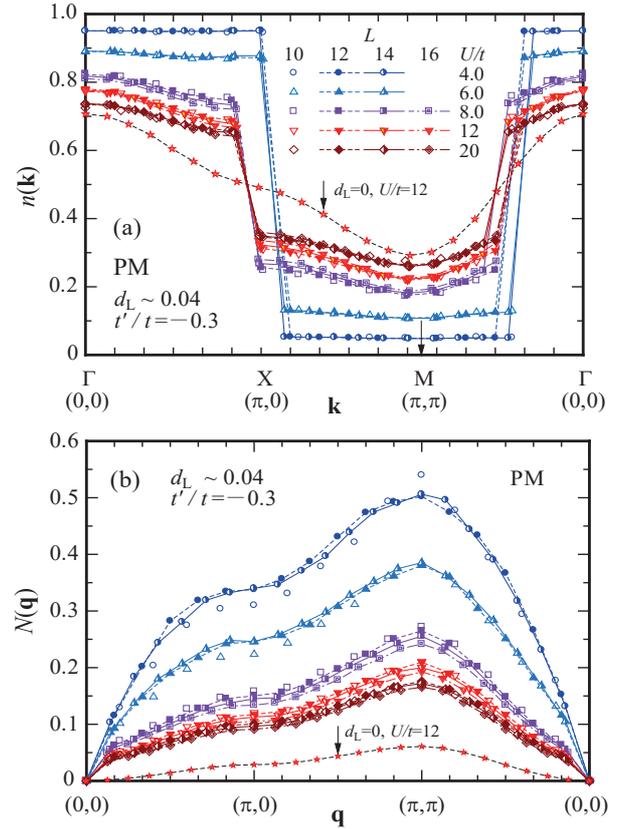}
\end{center}
\vskip -0.4cm
\caption{(Color online) 
(a) Momentum distribution function and (b) charge-density structure factor
in PM state with moderate value of $d_{\rm L}$ $\sim 0.04$ for several 
values of $U/t$ and $L$ along path of 
${\bf k}$ (${\bf q}$): $(0,0)$--$(\pi,0)$--$(\pi,\pi)$--$(0,0)$. 
For comparison, data for $\Psi_0$ ($d_{\rm L}=0$ and $U/t=12$) are also 
plotted. 
Symbols and colors are common in the two panels.
}
\label{fig:nk-nq-PM}
\end{figure}

%==========================================
Let us start with the behavior of the lowest-energy state $\Psi_0$. 
In Fig.~\ref{fig:d-vs-u-PM}, a sign of the Mott transition is found as 
anomaly in $d$ for $d_{\rm L}=0$ at $U_{\rm c}/t\sim 8.5$ indicated 
by an arrow. 
Whether this indicates a Mott transition or not is confirmed by the 
behavior of momentum distribution function,   
\begin{equation}
n({\bf k})=
\frac{1}{2}\sum_\sigma\langle c^\dag_{\bf k\sigma}c_{\bf k\sigma}\rangle, 
\label{eq:nk}
\end{equation}
and charge density structure factor, 
\begin{equation}
N({\bf q})=\frac{1}{N_{\rm s}} 
\sum_{i,j}e^{i{\bf q}\cdot({\bf R}_i-{\bf R}_j)} 
\left\langle{n_{i} n_{j}}\right\rangle - n^2. 
\label{eq:nq}
\end{equation} 
A previous study\cite{YOT} which treated $\Psi_0$ showed that a 
discontinuity at ${\bf k}={\bf k}_{\rm F}$ in $n({\bf k})$, namely, 
Fermi surface (FS) disappears for $U>U_{\rm c}$,\cite{note-YOT} and 
simultaneously $N({\bf q})$ becomes quadratic-like for small 
$|{\bf q}|$,\cite{note-N(q),Auerbach} as shown in Fig.~\ref{fig:nk-nq-PM} 
with small red stars for comparison, indicating that a gap opens in the 
charge sector. 
\par

Now, we consider how this Mott transition varies if extra doublons are 
created by introducing $d_{\rm L}$.  
In Fig.~\ref{fig:d-vs-u-PM}, there is no anomaly found in $d$ at $U\sim W$ 
for $d_{\rm L}>0$ even for large $L$ (see for $d_{\rm L}\sim 0.04$), 
suggesting that the Mott transition vanishes for $d_{\rm L}>0$. 
To corroborate it, in Fig.~\ref{fig:nk-nq-PM}, $n({\bf k})$ and 
$N({\bf q})$ are shown for $d_{\rm L}\sim 0.04$. 
In contrast to the case of $d_{\rm L}=0$ [see also Figs.~17 and 19 in 
Ref.~\citen{YOT}], $n({\bf k})$ exhibits discontinuities 
at near $(\pi,0)$ and $(\pi/2,\pi/2)$ and the behavior of $N({\bf q})$ 
for $|{\bf q}|\rightarrow 0$ is linear, both even for $U\gtrsim W$. 
Thus, the Mott transition vanishes at least for $d_{\rm L}\sim 0.04$.
\par

\begin{figure}[t]
%\vskip 0.2cm
%\hskip -0.2cm
\begin{center}
\includegraphics[width=8.5cm]{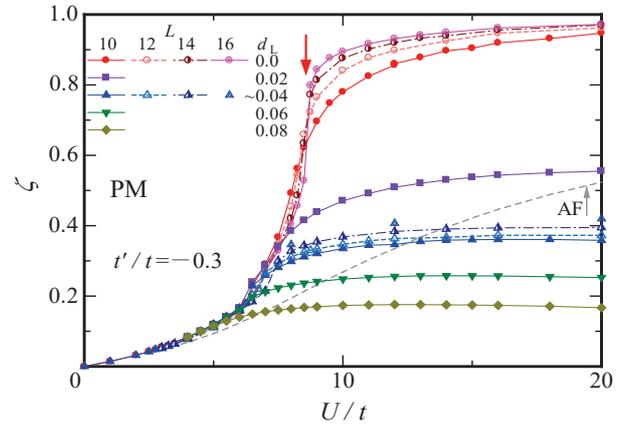}
\end{center}
\vskip -0.5cm
\caption{(Color online) 
$U/t$ dependence of optimized doublon-holon binding parameter in 
$\Psi_{d_{\rm L}}^{({\rm PM})}$ for several values of excitation 
intensity $d_{\rm L}$.
The Mott transition point for $d_{\rm L}=0$ is indicated by a red arrow. 
For later comparison (in Sect.~\ref{sec:AF}), the optimized $\zeta$ in the 
AF state for $d_{\rm L}=0$ and $L=12$ is added with a gray dashed curve. 
As for $d$-SC, the behavior of $\zeta$ is essentially similar to the 
present case of $\Psi_{d_{\rm L}}^{({\rm PM})}$.
}
\label{fig:para-zeta}
\end{figure} 
\begin{figure}[t]
%\vskip -0.5cm
%\hskip -0.2cm
\begin{center}
\includegraphics[width=8.5cm]{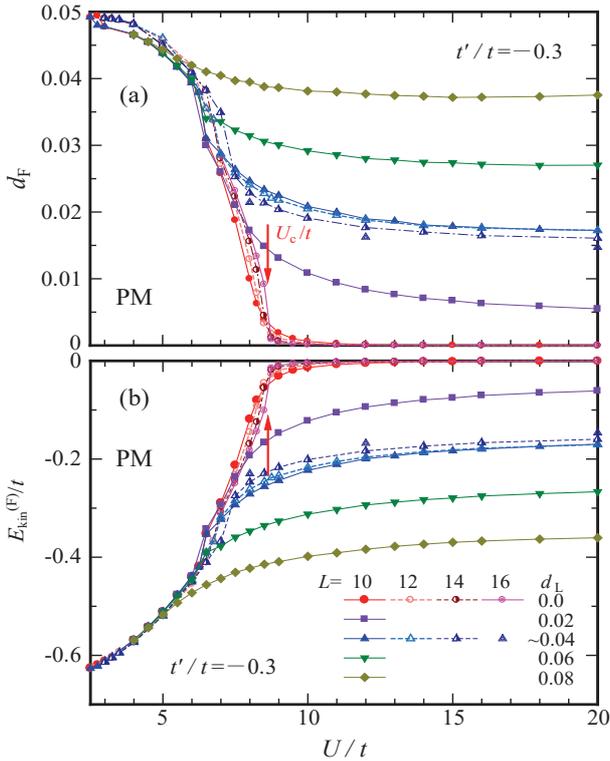}
\end{center}
\vskip -0.4cm
\caption{(Color online) 
$U/t$ dependence of (a) free doublon density and (b) part of kinetic 
energy $E_{\rm kin}^{({\rm F})}$, which stems from transfer of free 
doublons, for several values of $d_{\rm L}$ and $L$. 
The red arrows indicate the Mott transition point for $d_{\rm L}=0$. 
Symbols and colors are common in the two panels. 
}
\label{fig:df-Ei-vs-u-PM}
\end{figure} 

To consider the conductivity for $U>U_{\rm c}$ in the PM and $d$-SC states, 
it is convenient to introduce a notion of ``free doublon".
In Mott insulators ($d_{\rm L}=0$ and $U>U_{\rm c}$), almost all doublons 
are paired with holons (D-H bound pairs), and there is no free charge carrier 
(unpaired doublon or holon). 
This point is confirmed by the behavior of the D-H binding parameter $\zeta$ 
[Eq.~(\ref{eq:SymQ})] shown in Fig.~\ref{fig:para-zeta}. 
For $d_{\rm L}=0$, $\zeta$ approaches $1$ for $U>U_{\rm c}$ as $L$ increases, 
indicating that D-H pairs are tightly bound. 
However, as soon as $d_{\rm L}$ is introduced, $\zeta$ rapidly decreases and 
the D-H binding becomes weaker.\cite{note-binding} 
Consequently, isolated doublons and holons will appear, which we call free 
doublons (holons). 
In the present wave function, a doublon without a holon(s) in the 
nearest-neighbor sites is regarded as a free doublon, because the present 
D-H binding factor ${\cal P}_Q$ ranges over only the nearest-neighbor sites. 
Henceforth, the expectation value of the total number (density) of such 
free doublons is denoted by $D_{\rm F}$ ($d_{\rm F}=D_{\rm F}/N_{\rm s}$).
In Fig.~\ref{fig:df-Ei-vs-u-PM}(a), we show $U/t$ dependence of 
$d_{\rm F}$ for various values of $d_{\rm L}$. 
For $d_{\rm L}=0$, $d_{\rm F}$ becomes substantially zero for 
$U>U_{\rm c}$.\cite{note-dF}
Namely, almost all doublons (recall Fig.~\ref{fig:d-vs-u-PM} and 
Table \ref{table:BR}) exist as bound neutral D-H pairs. 
When $d_{\rm L}$ is raised to finite, free doublons come to survive for 
$U>U_{\rm c}$, suggesting that these free doublons contribute to 
conductivity.  
\par 

We can confirm this point by analyzing the kinetic energy 
$E_{\rm kin}=\langle{\mathscr H}_{\rm kin}\rangle$ 
into two contributions: 
\begin{equation}
E_{\rm kin}=E_{\rm kin}^{({\rm L})}+E_{\rm kin}^{({\rm F})},  
\label{eq:Ekin-ana}
\end{equation}
where $E_{\rm kin}^{({\rm L})}$ [$E_{\rm kin}^{({\rm F})}$] is the 
contribution from electron hopping that changes [preserves] 
${\cal D}$.\cite{Tocchio,YOTKT} 
$E_{\rm kin}^{({\rm L})}$ corresponds to the local process in which 
a D-H pair is created or destroyed and does not contribute to 
conductivity, whereas $E_{\rm kin}^{({\rm F})}$ chiefly consists of 
the global motion of free doublons or holons. 
In Fig.~\ref{fig:df-Ei-vs-u-PM}(b), $E_{\rm kin}^{({\rm F})}$ is shown 
as a function of $U/t$. 
One may notice that the behavior of $\left|E_{\rm kin}^{({\rm F})}\right|$ 
is quite similar to that of free-doublon density 
[Fig.~\ref{fig:df-Ei-vs-u-PM}(a)], indicating that free doublons and 
holons are the charge carriers.\cite{note-FD} 
\par

\begin{figure}[t]
\vskip -0.2cm
%\hskip -0.2cm
\begin{center}
\includegraphics[width=8.5cm]{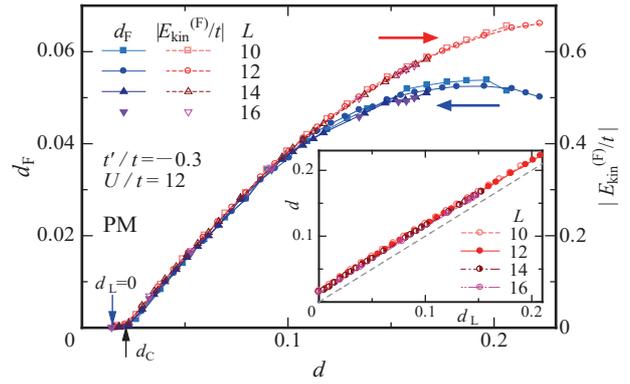}
\end{center}
\vskip -0.5cm
\caption{(Color online) 
Free-doublon density (left axis) and its contribution to $E_{\rm kin}/t$ 
(right axis) as functions of doublon density. 
The blue arrow on the horizontal axis indicates the value of $d$ in the 
lowest-energy state: $d(0)$. 
The black arrow indicates the threshold value of metallization $d_{\rm c}$. 
The inset shows doublon density as a function of the given lowest doublon 
density. 
The gray dashed line shows a guide line of $d=d_{\rm L}$
}
\label{fig:dF-vs-d-JPSJ}
\end{figure} 
We next discuss how $d_{\rm F}$ behaves as a function of $d_{\rm L}$ or 
rather $d$ for $U>U_{\rm c}$; various quantities including $d_{\rm F}$ 
better scale to $d$ than to $d_{\rm L}$. 
The relation between $d$ and $d_{\rm L}$ is shown for a typical value 
$U/t=12$ in the inset of Fig.~\ref{fig:dF-vs-d-JPSJ}, and is broadly 
written as 
\begin{equation}
d=d_{\rm L}+d(0). 
\label{eq:d-dL}
\end{equation}
This relation is also read from Fig.~\ref{fig:d-diff} and available 
for the $d$-SC and AF states for large $U/t$. 
Thus, we often use $d$ instead of $d_{\rm L}$. 
In the main panel of Fig.~\ref{fig:dF-vs-d-JPSJ}, $d$ dependence of 
$d_{\rm F}$ and $\left|E_{\rm kin}^{({\rm F})}\right|/t$ is shown. 
In weakly excited cases ($d\lesssim 0.1$, $d_{\rm L}\lesssim 0.08$), 
$d_{\rm F}$ and $\left|E_{\rm kin}^{({\rm F})}\right|$ are proportional 
to each other. 
In strongly excited cases ($d\gtrsim 0.1$, $d_{\rm L}\gtrsim 0.08$), 
however, the behavior of $\left|E_{\rm kin}^{({\rm F})}\right|$ deviates 
from $d_{\rm F}$, because hopping of a bound doublon between D-H 
pairs (clusters) comes to occur frequently as $d$ increases and does not 
change $d_{\rm F}$ but contributes to $E_{\rm kin}^{({\rm F})}$. 
\par

In Fig.~\ref{fig:dF-vs-d-JPSJ}, we find a narrow but finite range of 
$d_{\rm F}=0$ for $d(0)<d<d_{\rm c}\sim 0.02$ ($0<d_{\rm L}\lesssim 0.01$), 
where $\Psi^{({\rm PM})}_{d_{\rm L}}$ is insulating. 
Therefore, we need finite excitation intensity $d_{\rm c}$ to metallize 
$\Psi^{({\rm PM})}_{d_{\rm L}}$. 
This threshold $d_{\rm c}$ is somewhat larger for 
$\Psi^{(d{\rm -SC})}_{d_{\rm L}}$ as previously discussed [see Fig.~3(a) 
in Ref.~\citen{proc}]. 
The $d$-SC state remains insulating for $d_{\rm L}\lesssim 0.022$.
\par

%==========================================
\section{Superconducting States\label{sec:SC}}
\begin{figure}[t]
%\vskip 0.2cm
%\hskip -0.2cm
\begin{center}
\includegraphics[width=8.5cm]{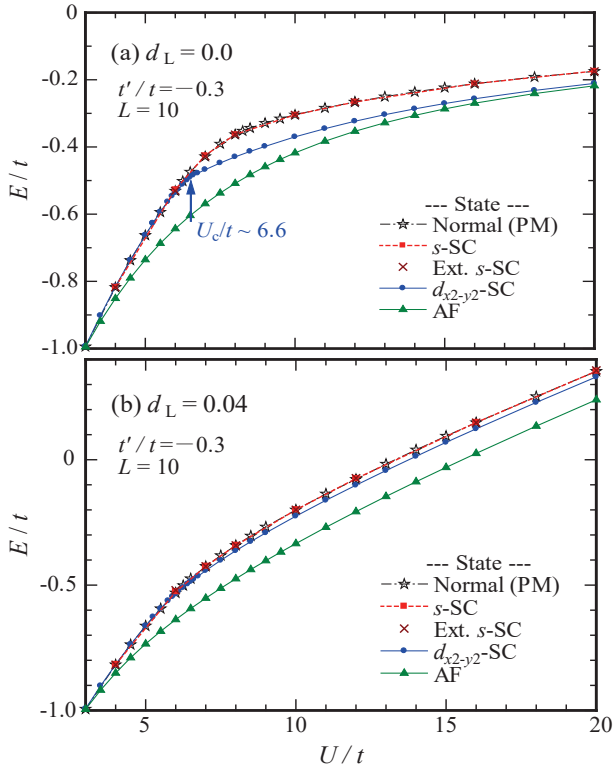}
\end{center}
\vskip -0.5cm
\caption{(Color online) 
Comparison of total energy per site among states treated here as function 
of correlation strength for (a) lowest-energy state, and (b) moderate 
excited state ($d_{\rm L}=4$). 
In (a), the Mott-transition point of $d$-SC state is indicated by a blue 
arrow. 
For clarity, data only for $L=10$ are plotted.  
}
\label{fig:E-comp}
\end{figure}

%==========================================
We first discuss the stability among the three SC states of different 
pairing symmetries. 
Previous studies for the lowest-energy state \cite{YS-SC,Gros,YO} 
in strongly correlated regimes showed that the $s$-wave and extended 
$s$-wave SC states bring about no energy reduction, namely, 
the optimized $\Psi_0^{(s-{\rm SC})}$ and 
$\Psi_0^{({\rm x-SC})}$ [Eq.~(\ref{eq:Phi_d})] are 
reduced to $\Psi_0^{({\rm PM})}$. 
On the other hand, $\Psi_0^{(d{\rm -SC})}$ exhibits appreciable energy 
reduction for $U/t\gtrsim 5$.\cite{YOTKT} 
These results are confirmed in Fig.~\ref{fig:E-comp}(a), where the total 
energies for $d_{\rm L}=0$ are compared among the five states studied 
here as functions of $U/t$. 
As shown in Fig.~\ref{fig:E-comp}(b), such situation does not qualitatively 
change in excited states of moderate $d_{\rm L}$, although the energy 
gain owing to $\Psi^{(d{\rm -SC})}_{d_{\rm L}}$ appreciably decreases. 
Because we found no energy gain owing to $\Psi_{d_{\rm L}}^{(s-{\rm SC})}$ 
and $\Psi_{d_{\rm L}}^{({\rm x-SC})}$ ($\Delta_s$, $\Delta_{\rm x}=0$) 
for any parameter set we studied, 
we will concentrate on the $d$-SC state in the following. 
We leave comparison with the AF state for Sect.~\ref{sec:AF}.
\par

A previous study\cite{YTOT,YOT} showed that the lowest-energy state 
$\Psi_0^{(d{\rm -SC})}$ exhibits a Mott transition at 
$U=U_{\rm c}\sim 6.6t$ [arrow in Fig.~\ref{fig:E-comp}(a)]. 
Therefore, the excited states with extra doublons are meaningful 
for $U>U_{\rm c}$, where $\Delta d\sim d_{\rm L}$ as shown 
in Fig.~\ref{fig:d-diff}(b). 
In this regime, various properties discussed for 
$\Psi_{d_{\rm L}}^{(\rm PM)}$ in Sect.~\ref{sec:PM} applies to 
$\Psi_{d_{\rm L}}^{(d{\rm -SC})}$. 
The main point is that for $d_{\rm L}>d_{\rm Lc}$ 
[$d_{\rm Lc}\sim 0.15~(0.22)$ for $U/t=8$~($12$)], 
$\Psi_{d_{\rm L}}^{(d{\rm -SC})}$ becomes superconducting, whose charge 
carriers are free (unbound) doublons and holons ($d_{\rm F}>0$). 
Since this point was discussed in the preceding article,\cite{proc} 
here we consider what was not taken up there, especially, $d$ 
($d_{\rm L}$) dependence. 
\par
\begin{figure}[t]
%\vskip 0.2cm
%\hskip -0.2cm
\begin{center}
\includegraphics[width=8.5cm]{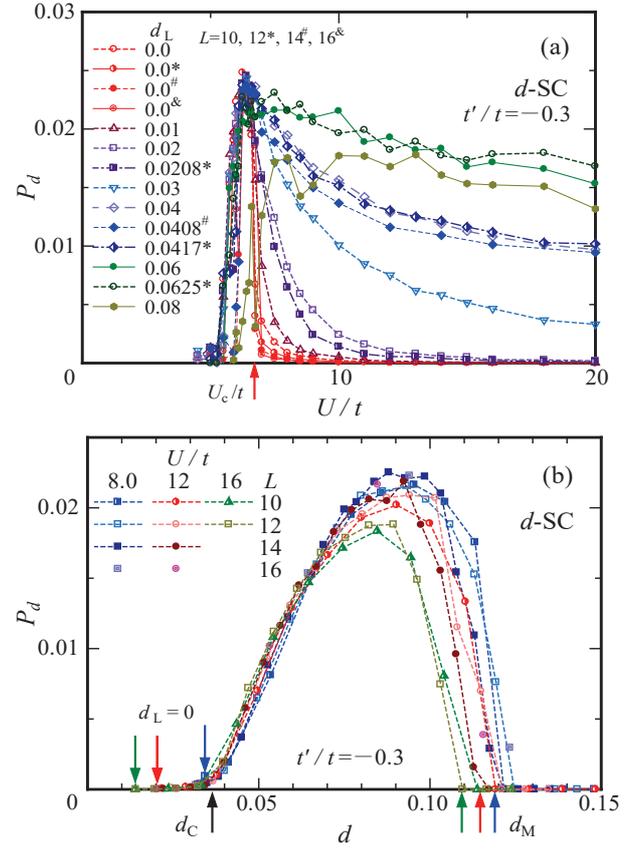}
\end{center}
\vskip -0.5cm
\caption{(Color online) 
$d_{x^2-y^2}$-wave SC correlation function as function of (a) interaction 
strength for various values of $d_{\rm L}$ and of (b) doublon density
for $U/t=8$, $12$, and $16$.
In (a), the Mott-transition point is indicated by a red arrow. 
In (b), some relevant values of $d$ are indicated by arrows on the 
horizontal axis.
}
\label{fig:Pd}
\end{figure} 

As a quantity to represent the strength of $d_{x^2-y^2}$-wave SC, we use 
the $d$-wave nearest-neighbor pair correlation function 
$P_d({\bf r})$  defined as,
%----------------------------------------------------------------------
\begin{align}
P_d({\bf r})=&\frac{1}{N_{\rm s}}
\sum_{i}\sum_{\tau,\tau'=\hat {\bf x},\hat {\bf y}}
(-1)^{1-\delta_{\tau,\tau'}}
%\times\qquad \nonumber\\ 
 \left\langle{\Delta _\tau^\dag({\bf R}_i)\Delta_{\tau'}
({\bf R}_i+{\bf r})}\right\rangle, 
\label{eq:pd}
\end{align}
%----------------------------------------------------------------------
in which $\hat{\bf x}$ and $\hat{\bf y}$ denote the unit vectors 
in the $x$ and $y$ directions, respectively, and 
$\Delta_\tau^\dag({\bf R}_i)$ is the creation operator of a
nearest-neighbor singlet, 
%----------------------------------------------------------------------
\begin{equation}
\Delta_\tau^\dag({\bf R}_i)=
(c_{{i}\uparrow}^\dag c_{{i}+\tau\downarrow}^\dag+ 
 c_{{i}+\tau\uparrow}^\dag c_{{i}\downarrow}^\dag)
 /{\sqrt 2}. 
\end{equation}
%----------------------------------------------------------------------
%
Because $P_d({\bf r})$ rapidly decays with $|{\bf r}|$ and has almost 
constant values for $|{\bf r}|\ge 3$ in the strongly correlated regimes 
(we actually checked it), we use, for accuracy, the average of $P_d({\bf r})$ 
with $|{\bf r}|\ge 3$ as a typical value $P_d$.\cite{note-Pd} 
To measure the distance $|{\bf r}|$, we use so-called the Manhattan 
(stepwise) metric. 
For details of $P_d({\bf r})$, see Ref.~\citen{YOTKT}. 
\par
\begin{figure}[t]
%\vskip 0.2cm
%\hskip -0.2cm
\begin{center}
\includegraphics[width=8.5cm]{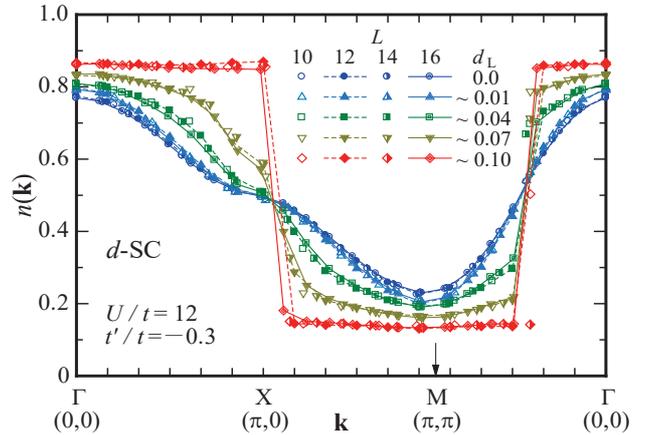}
\end{center}
\vskip -0.3cm
\caption{(Color online) 
Momentum distribution function in $d$-SC state along same path as in 
Fig.~\ref{fig:nk-nq-PM}(a) for various values of excitation intensity 
$d_{\rm L}$ for $U/t=12$ .
}
\label{fig:nk-d}
\end{figure} 
%\vskip -0.3cm
%
Figure \ref{fig:Pd}(a) shows $P_d$ as a function of $U/t$ for various 
values of excitation intensity $d_{\rm L}$.
For $d_{\rm L}=0$, we know that $P_d$ vanishes for $U>U_{\rm c}$.\cite{YOT}  
For $0<d_{\rm L}<d_{\rm Lc}$ ($0<d<d_{\rm c}\sim 0.037$),\cite{note:d-dL} 
$P_d$ is still substantially null for large $U/t$, as mentioned. 
In this range of $d_{\rm L}$, $\Psi_{d_{\rm L}}^{(d{\rm -SC})}$ remains 
insulating as shown in Fig.~\ref{fig:nk-d}, in which $n({\bf k})$ for 
$d_{\rm L}\sim 0.01$ (light blue) has no discontinuity on the path of 
${\bf k}$ [even near $(\pi/2,\pi/2)$]. 
For $d_{\rm L}>d_{\rm Lc}$, $P_d$ increases as $d_{\rm L}$ increases 
with a tail toward large $U/t$, becomes maximum 
at $d_{\rm L}\sim 0.06$ ($d\sim 0.095$), and then decreases. 
This behavior of $P_d$ becomes intelligible by plotting it as a function of 
$d$ as shown in Fig.~\ref{fig:Pd}(b) for $U/t=8$, $12$ and $16$. 
The formation of $d$-SC order in this range of $d_{\rm L}$ is corroborated 
by the behavior of $n({\bf k})$ in Fig.~\ref{fig:nk-d}; 
for $d_{\rm L}\sim 0.04$ and $\sim 0.07$, $n({\bf k})$ has no discontinuity 
near the antinodal point ${\bf k}=(\pi,0)$ but exhibits a discontinuity 
(FS) near $(\pi/2,\pi/2)$ in the nodal direction.\cite{Paramekanti} 
For $d\gtrsim 0.1$, $P_d$ rapidly drops and vanishes again at 
$d=d_{\rm M}\sim 0.11$--$0.12$. 
$\Psi_{d_{\rm L}}^{(d{\rm -SC})}$ becomes metallic for $d>d_{\rm M}$, 
where $n({\bf k})$ exhibits 
discontinuities both near $(\pi,0)$ and $(\pi/2,\pi/2)$ as seen for 
$d_{\rm L}\sim 0.10$ ($d\sim 0.12$) in Fig.~\ref{fig:nk-d}. 
Thus, within $\Psi_{d_{\rm L}}^{(d{\rm -SC})}$, a $d$-SC order forms 
for moderate excitation intensity $0.02\lesssim d_{\rm L}\lesssim 0.11$. 
\par

Note that the maximal value of $P_d$ for a fixed value of $U/t$ 
[Fig.~\ref{fig:Pd}(a)] is broadly equal to or slightly smaller than 
the corresponding value in chemically doped case [Fig.~24(b) in 
Ref.~\citen{YOTKT}]. 
Furthermore, the maximum of $P_d$ for $d_{\rm L}>0$ never becomes greater 
than that for $d_{\rm L}=0$ at $U\sim U_{\rm c}$. 
These results suggest that higher $T_{\rm c}$ is not attained by D-H 
excitation at half filling, even if $\Psi_{d_{\rm L}}^{(d{\rm -SC})}$ 
is realized. 
\par 

\begin{figure}[t]
%\vskip 0.2cm
%\hskip -0.2cm
\begin{center}
\includegraphics[width=8.5cm]{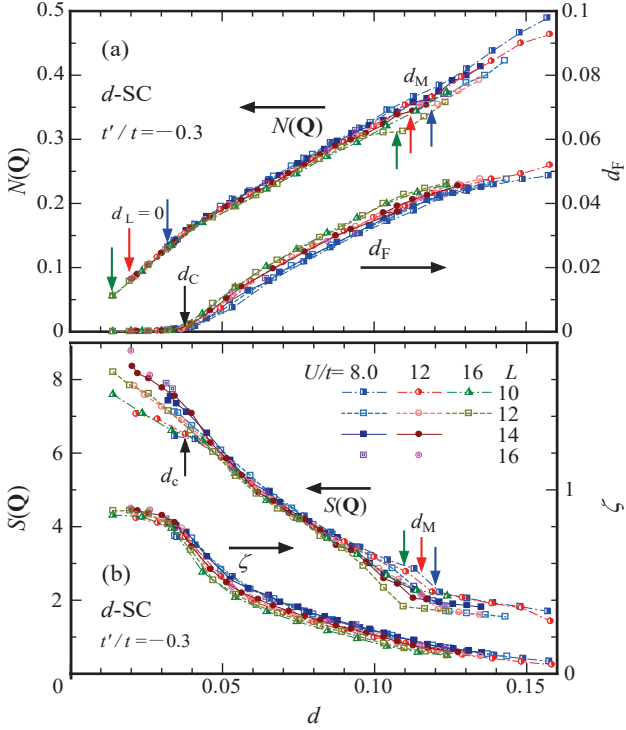} 
\end{center}
\vskip -0.5cm
\caption{(Color online) 
(a) Charge-density structure factor $N({\bf q})$ at 
${\bf q}={\bf Q}=(\pi,\pi)$ and free doublon density, and (b) spin structure 
factor $S({\bf Q})$ and optimized D-H binding parameter, as functions of 
doublon density for a few values of $U/t$. 
Symbols and lines are common in the two panels. 
The arrows of blue, red, and green for $d_{\rm L}=0$ and $d_{\rm M}$ 
indicate the data points of $U/t=8$, $12$, and $16$, respectively. 
}
\label{fig:nq-sq-dSC}
\end{figure} 
Finally, we consider how charge and spin correlations evolves as excitation 
intensity is varied. 
In Fig.~\ref{fig:nq-sq-dSC}, we show $d$ dependence of the charge-density 
structure factor $N({\bf q})$ [Eq.~(\ref{eq:nq})] and spin structure factor
\begin{equation} 
S({\bf q})=\frac{1}{N_{\rm s}}\sum_{ij}{e^{i{\bf q}
\cdot({\bf R}_i-{\bf R}_j)} 
\left\langle{S_{i}^zS_{j}^z}\right\rangle}, 
\label{eq:sq}
\end{equation} 
at ${\bf q}={\bf Q}=(\pi,\pi)$, where both $N({\bf q})$ and $S({\bf q})$ 
become maximum. 
The charge and spin correlations are scaled by doublon density rather 
than $U/t$.  
Generally, if repulsive electron correlation becomes stronger, $N({\bf q})$ 
[$S({\bf q})$] decreases [increases]. 
Therefore, Fig.~\ref{fig:nq-sq-dSC} means that the D-H-pair excitation 
greatly weakens the effective electron correlation in the system; 
$d_{\rm F}$ increases and the D-H binding becomes loose ($\zeta$ decreases). 
Because the $d$-SC correlation in the present system is broadly given 
by the product of charge and spin correlations,\cite{ZGRS,YOTKT} 
The behavior of $P_d$ is mainly controlled by $N({\bf q})$ or $d_{\rm F}$ 
shown in Fig.~\ref{fig:nq-sq-dSC}(a) in the vicinity of $d_{\rm c}$, but by 
the decay of AF spin correlation, which causes the $d$-SC 
pairing,\cite{Scalapino} near $d_{\rm M}$. 
\par

%==========================================
\section{Antiferromagnetic States\label{sec:AF}}
\begin{figure}[t]
%\vskip 0.2cm
%\hskip -0.2cm
\begin{center}
\includegraphics[width=8.5cm]{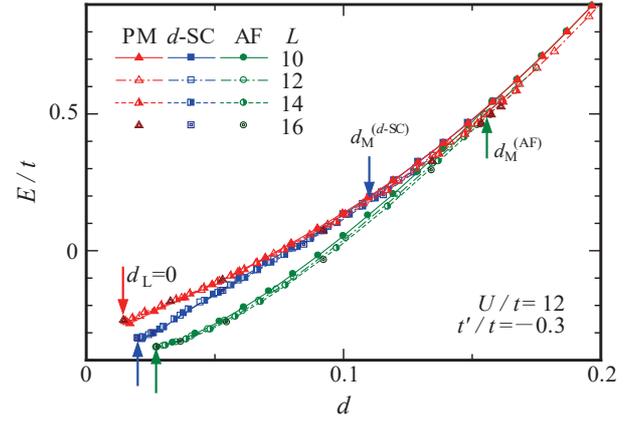}
\end{center}
\vskip -0.3cm
\caption{(Color online) 
Comparison of total energy per site as functions of doublon density 
among $\Psi_{\rm PM}$, $\Psi_d$, and $\Psi_{\rm AF}$ for $U/t=12$. 
The boundary points $d_{\rm M}$ of the $d$-SC and AF orders are indicated 
by arrows. 
The situation is qualitatively the same for another $U/t$ ($>U_{\rm c}$). 
}
\label{fig:Etot-vs-d}
\end{figure}

%==========================================
To begin with, we study the stability of $\Psi_{d_{\rm L}}^{({\rm AF})}$. 
Figure \ref{fig:E-comp}(b) shows that the AF state has the lowest energy 
also in an excited state ($d_{\rm L}=0.04$) among the states treated here
for $U>U_{\rm AF}$. 
In Fig.~\ref{fig:Etot-vs-d}, $E/t$ is compared as functions of $d$ for 
$U/t=12$ [Eq.~(\ref{eq:d-dL}) is roughly valid here]; 
$\Psi_{d_{\rm L}}^{({\rm AF})}$ is 
still stable as long as $m$ is finite ($d_{\rm L}<d_{\rm LM}\sim 0.14$). 
Thus, the AF state always overcomes the $d$-SC state in the 
$d_{\rm L}$--$U/t$ space, and will be realized as a stationary state. 
This aspect is consistent with the $\mathscr{D}=1$ case (corresponding to 
$d_{\rm L}=0.0625$) of the previous study\cite{Takahashi1} for the 
$t$-$J$-type model. 
\par 

\begin{figure}[t]
%\vskip 0.2cm
%\hskip -0.2cm
\begin{center}
\includegraphics[width=8.5cm]{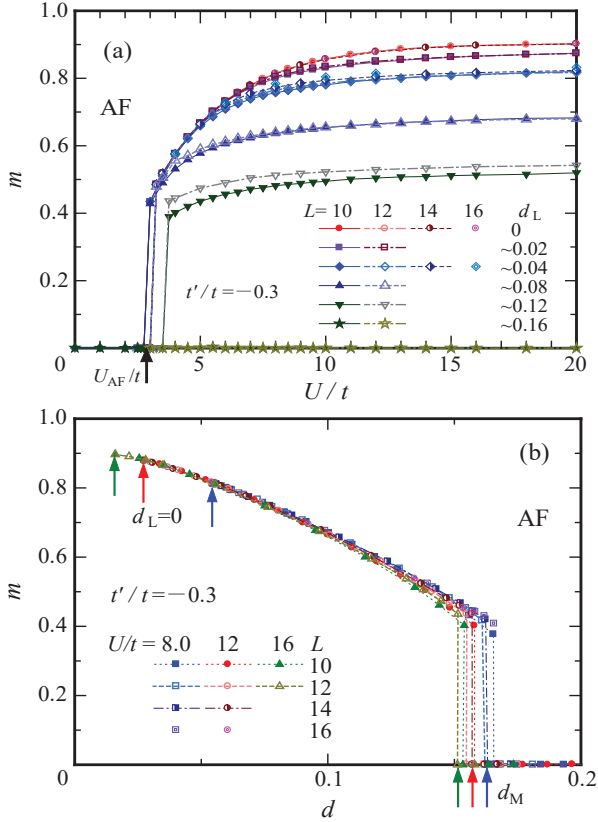}
\end{center}
\vskip -0.3cm
\caption{(Color online) 
Sublattice magnetization of $\Psi_{\rm AF}$, (a) as function of $U/t$
for various values of excitation intensity $d_{\rm L}$ and $L$, and (b) 
as function of $d$ for three values of $U/t$. 
The AF transition points ($U_{\rm AF}/t$, $d_{\rm M}$) are indicated 
respectively by arrows on the horizontal axes. 
}
\label{fig:mag-L} 
\end{figure} 
Therefore, it is important to study relevant properties of the AF state. 
The nature of AF gradually changes from a Slater type to a Mott type 
around $U=W=8t$ in the lowest-energy state $\Psi_0^{({\rm AF})}$.\cite{TY} 
This crossover seems to be preserved in the excited states as seen 
in Fig.~\ref{fig:d-diff}(c), where $\Delta d$ smoothly converges to the 
limiting values of $U/t\rightarrow\infty$.
Thus, the present D-H excitation scheme is considered to be appropriate 
for $U\gtrsim W$ (Mott regime).  
\par

In Fig.~\ref{fig:mag-L}(a), we show $U/t$ dependence of the sublattice 
magnetization (an order parameter of AF), 
\begin{equation}
m=\frac{2}{N_{\rm s}}\sum_j
\left| e^{i{\bf Q}\cdot{\bf r}_j}\langle S^{z}_j\rangle \right| 
\qquad \mbox{with}~~{\bf Q}=(\pi,\pi), 
\label{eq:mag}
\end{equation}
which becomes $1$ at the full moment. 
For $t'/t=-0.3$, the optimized state is paramagnetic for 
$U<U_{\rm AF}\sim 3t$. 
At $U=U_{\rm AF}$, $\Psi_{d_{\rm L}}^{({\rm AF})}$ exhibits a first-order 
AF transition and $m$ discontinuously appears and gradually increases 
for $U>U_{\rm AF}$. 
The transition value $U_{\rm AF}/t$ is almost independent of $d_{\rm L}$, 
but becomes somewhat larger as $d_{\rm L}$ approaches the vanishing point 
$d_{\rm LM}\sim 0.14$. 
In Fig.~\ref{fig:mag-L}(b), $d$ dependence of $m$ is shown; as $d$ 
($d_{\rm L}$) increases toward $d_{\rm M}$ ($d_{\rm LM}$), $m$ monotonically 
decreases and finally vanishes at a first-order transition point 
$d_{\rm M}\sim 0.16$. 
\par

\begin{figure}[t]
%\vskip 0.2cm
%\hskip -0.2cm
\begin{center}
\includegraphics[width=8.5cm]{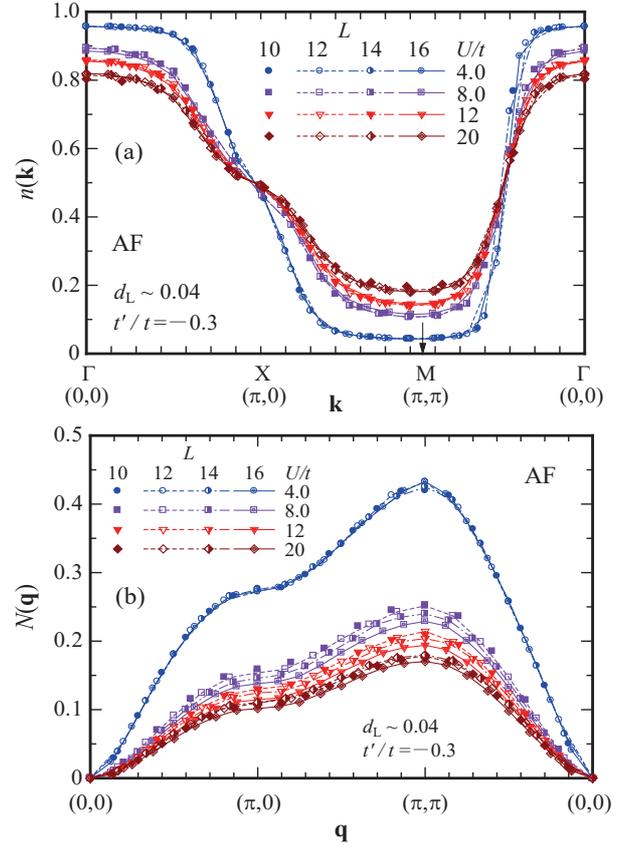}
\end{center}
\vskip -0.3cm
\caption{(Color online) 
(a) Momentum distribution function and charge-density structure factor 
for AF state with $d_{\rm L}\sim 0.04$ for four values of $U/t$ and $L$ 
along same path as in Fig.~\ref{fig:nk-nq-PM}.
}
\label{fig:nk-nq-vs-u-AF}
\end{figure} 
\begin{figure}[t]
\vskip -0.2cm
%\hskip -0.2cm
\begin{center}
\includegraphics[width=8.5cm]{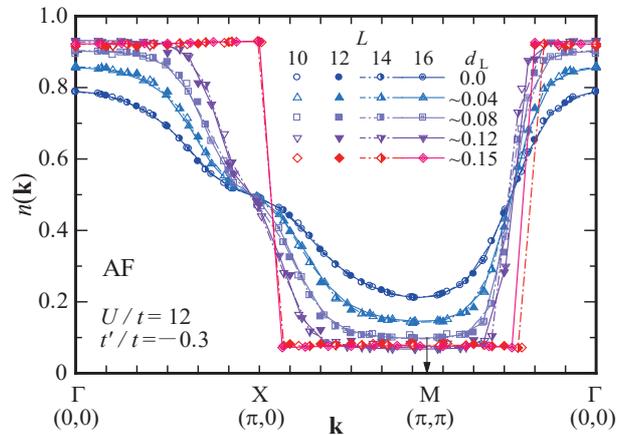}
\end{center}
\vskip -0.3cm
\caption{(Color online) 
Momentum distribution function of $\Psi_{\rm AF}$ for $U/t=12$. 
The excitation intensity $d_{\rm L}$ is varied. 
}
\label{fig:nk-dL-af}
\end{figure} 

Now, we consider the conductivity of $\Psi_{d_{\rm L}}^{({\rm AF})}$ 
($d_0<d<d_{\rm M}$) and $\Psi_{d_{\rm L}}^{({\rm PM})}$ ($d>d_{\rm M}$). 
Figure \ref{fig:nk-nq-vs-u-AF} shows $n({\bf k})$ and $N({\bf q})$ of 
$\Psi_{d_{\rm L}}^{({\rm AF})}$ with $d_{\rm L}\sim 0.04$ for four values 
of $U/t$. 
In contrast to the PM (Fig.~\ref{fig:nk-nq-PM}) and $d$-SC [Figs.~4(b) 
and 5(b) in Ref.~\citen{proc}] states, which are always (super)conducting 
for $d\sim 0.04$, the AF is always insulating. 
Namely, $\Psi_{d_{\rm L}}^{({\rm AF})}$ exhibits no Fermi surface 
[discontinuity in $n({\bf k})$] and is gapped [downward convex behavior 
of $N({\bf q})$ for $|{\bf q}|\rightarrow 0$]. 
In Fig.~\ref{fig:nk-dL-af}, we show the evolution of $n({\bf k})$ as 
$d_{\rm L}$ increases for $U/t=12$. 
As far as $m$ is finite ($d_{\rm L}<d_{\rm LM}\sim 0.14$), the gap remains. 
A Fermi surface appears for $d_{\rm L}>d_{\rm LM}$. 
This behavior of the excited AF state (by introduction of D and H) is 
distinct from that of a partially filled AF state obtained by chemical 
doping with holons or doublons; the latter state is always metallic 
with pocket-type Fermi surfaces.\cite{YOTKT} 
A fundamental difference between the two cases is whether the charge 
balance or neutrality is preserved (former) or not (latter) in the models. 
%In contrast, both PM and $d$-SC states are conductive in both cases. 
\par

\begin{figure}[t]
%\vskip 0.2cm
%\hskip -0.2cm
\begin{center}
\includegraphics[width=8.5cm]{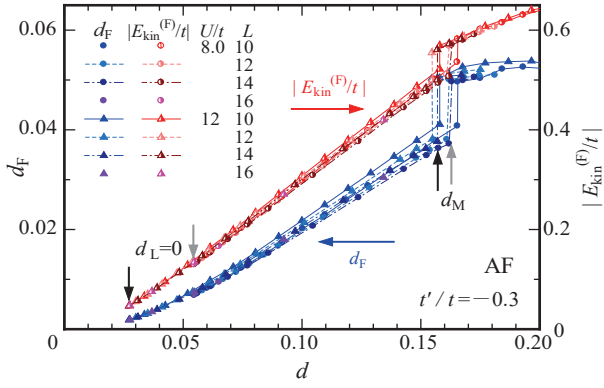}
\end{center}
\vskip -0.3cm
\caption{(Color online) 
Free-doublon density (cool colors, left axis) and absolute value of 
$E_{\rm kin}^{({\rm F})}/t$ (warm colors, right axis) in AF state 
as functions of doublon density.
The values of the lowest energy state ($d_{\rm L}=0$) and of transition
points ($d_{\rm M}$) are indicated by black ($U/t=12$) and gray ($U/t=8$)
arrows.
}
\label{fig:df-vs-d-af}
\end{figure} 

In this connection, we mention the role of free doublons in 
$\Psi_{d_{\rm L}}^{({\rm AF})}$. 
As discussed in Sects.~\ref{sec:PM} and \ref{sec:SC}, free doublons 
and holons are responsible for conductivity in 
$\Psi_{d_{\rm L}}^{({\rm PM})}$ and $\Psi_{d_{\rm L}}^{(d{\rm -SC})}$. 
In Fig.~\ref{fig:df-vs-d-af}, we plot $d_{\rm F}$ and 
$\left|E_{\rm kin}^{({\rm F})}\right|/t$ [in Eq.~(\ref{eq:Ekin-ana})] 
as functions of $d$. 
In contrast to the PM (Fig.~\ref{fig:dF-vs-d-JPSJ}) and $d$-SC 
[Fig.~\ref{fig:nq-sq-dSC}(a)] cases, free doublons and holons already 
exist in the lowest-energy state ($d_{\rm L}=0$) for $U/t=8$ and $12$, 
which is insulating. 
When doublons and holons are excited (as $d_{\rm L}$ or $d$ increases), 
free-doublon density increases; $d_{\rm F}$ for 
$\Psi_{d_{\rm L}}^{({\rm AF})}$ is only 
somewhat smaller than $d_{\rm F}$ for $\Psi_{d_{\rm L}}^{({\rm PM})}$ 
and $\Psi_{d_{\rm L}}^{(d{\rm -SC})}$. 
Nevertheless, the AF state is insulating as far as $m$ is finite 
($d<d_{\rm M}$). 
This means that the D-H binding mechanism plays at most a subsidiary 
role for the insulation of the AF state. 
Actually, the optimized value of D-H binding parameter $\zeta$ for 
$\Psi_{\rm AF}$ is much smaller than those for the other states 
(Fig.~\ref{fig:para-zeta}). 
We should pursue the leading factor for the insulation of 
$\Psi_{d_{\rm L}}^{({\rm AF})}$. 
\par

%==========================================
\section{Summary and Discussions\label{sec:summary}} 
%==========================================
Assuming that high-energy quasi-stationary states are generated by 
photoexcitation or tera-Hertz pulse electric field, etc., we statically 
studied excited states exceeding the Mott gap ($\Delta E\gtrsim U$) 
in the paramagnetic (normal), superconducting ($d_{x^2-y^2}$-wave, 
$s$-wave, and extended $s$-wave symmetries), and antiferromagnetic 
branches for the Hubbard model at half filling. 
We applied a variational Monte Carlo method to the excited states 
by imposing the minimum doublon densities $d_{\rm L}$ on the trial 
states; $d_{\rm L}$ was found to correspond to the excitation intensity 
such as optical intensity per Cu site for cuprates. 
We recapitulate the main results including brief discussions in the 
following. 
\par

(1) In the PM and SC cases, the states become conductive over the threshold 
of excitation intensity $d_{\rm Lc}$ [$\sim 0.01$ (PM), $\sim 0.02$ 
($d$-SC)].
In this regime ($d_{\rm L}>d_{\rm Lc}$), free doublons and holons generated 
in excitation become charge carriers.
\par

(2) In the same way as the lowest energy states, the SC states with 
$s$-wave and extended-$s$-wave symmetries have no energy gain over that 
of the PM state as excited states ($d_{\rm L}>0$) for any parameter set 
we checked. 
From this standpoint, an $\eta$-pairing state,\cite{Yang} which has an 
$s$-wave-type symmetry and may arises immediately after 
excitation,\cite{eta} is unlikely to be stable as a quasi-stationary state
after some energy dissipation. 
\par

(3) The $d$-SC state becomes more stable than the PM state for intermediate 
excitation intensity $d_{\rm Lc}\lesssim d_{\rm L}\lesssim 0.11$; 
as a function of $d_{\rm L}$, the pairing correlation function $P_d$ 
exhibits a maximum at $d_{\rm L}\sim 0.07$. 
The $U/t$ dependence of $P_d$ in $\Psi_{d_{\rm L}}^{(d{\rm -SC})}$ 
[Fig.~\ref{fig:Pd}(a)] is similar to the behavior of the chemically doped 
cases of the lowest-energy state $\Psi_0^{(d{\rm -SC})}$ [Fig.~24(b) in 
Ref.~\citen{YOTKT}]. 
The maximum of $P_d$ is subtly smaller for $\Psi_{d_{\rm L}}^{(d{\rm -SC})}$ 
for any fixed value of $U/t$ ($U>U_{\rm c}$). 
Therefore, the strength of superconductivity (or $T_{\rm c}$) induced 
by excitation at half filling is unlikely to exceed that obtained by doping 
holes to the lowest energy state (as usually done). 
\par

\begin{figure}[t]
%\vskip 0.2cm
%\hskip -0.2cm
\begin{center}
\includegraphics[width=7.0cm]{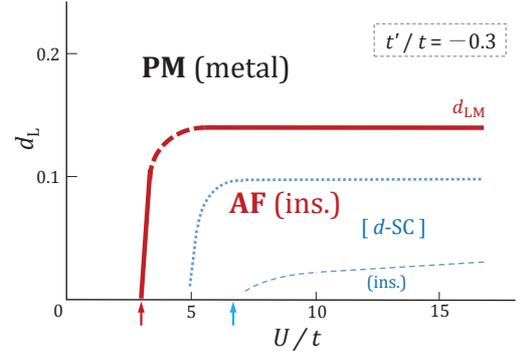}
\end{center}
\vskip -0.3cm
\caption{(Color online) 
Semi-quantitative phase diagram of stationary D-H excited state in 
space of excitation intensity ($d_{\rm L}$) and interaction strength 
($U/t$) constructed within present wave functions. 
The loci of the conductor-insulator transitions for the two orders at 
$d_{\rm L}=0$ ($U_{\rm AF}/t$, $U_{\rm c}/t$) are indicated by arrows 
of corresponding colors, respectively. 
The $d$-SC state is always metastable and appears if the AF order is 
destroyed by some reason.
}
\label{fig:phase-d}
\end{figure} 

(4) The AF state is the most stable among the states we treated as far as 
the order parameter ($m$) is finite ($0\le d_{\rm L}\lesssim 0.14$). 
Therefore, the AF state is reached after energy dissipation processes 
without changing the number of doublons. 
In contrast to the PM and $d$-SC states, the AF state preserves insulating
in the whole excited regime of AF order. 
As a summary, we construct a semi-quantitative phase diagram in the 
$U/t$--$d_{\rm L}$ space shown in Fig.~\ref{fig:phase-d}. 
There is no range where the $d$-SC order appears, which is always unstable 
toward the AF order. 
It follows that when experiments observe that the state is 
conductive,\cite{Okamoto,Miyamoto,Terashige} the optical intensity is 
stronger than $d_{\rm LM}$ or the probes may observe the behavior of 
transient states. 
\par

(5) The effect of $t'$ is qualitatively negligible for moderate $d_{\rm L}$
in all the states we addressed (see Appendix).   

Finally, we add a comment on the relation between the present study with 
the previous ones\cite{Takahashi1,Takahashi2,Gomi} for a $t$-$J$-type 
model. 
In the present study, the bases of ${\cal D}<D_{\rm L}$ are excluded, 
whereas the effects of ${\cal D}<{\mathscr D}-1$ and 
${\cal D}>{\mathscr D}+1$ are not included in the previous studies. 
Therefore, the results may somewhat differ except for 
$U/t\rightarrow\infty$. 
In the last paper\cite{Gomi} of this series, effects of repulsive 
interaction between nearest-neighbor sites are considered, which turns to 
attractive interaction between nearest-neighbor D-H pairs. 
As a result, a phase separation takes place, where clusters of doublons 
and holons alternately sitting separate from domains of singly occupied 
sites. 
This phase-separated state is also insulating with partial AF orders. 
A conductive excited state may arise from another
factor;\cite{future} 
we leave it for future studies. 
\par

%\begin{acknowledgment}
%\acknowledgment
\section*{Acknowledgment}
One of the authors (HY) thanks Philipp Werner and the late Sumio Ishihara 
for comments on an early stage of this study. 
This work is supported in part by Grant-in-Aids from the Ministry of 
Education, Culture, Sports, Science and Technology, Japan. 
\par
%\end{acknowledgment}

\appendix
\section{Effect of Diagonal Hopping\label{sec:diagonal}}
\begin{figure}[h]
\begin{center} 
%\vskip -5mm
%\hskip -20mm
\includegraphics[width=8.5cm,clip]{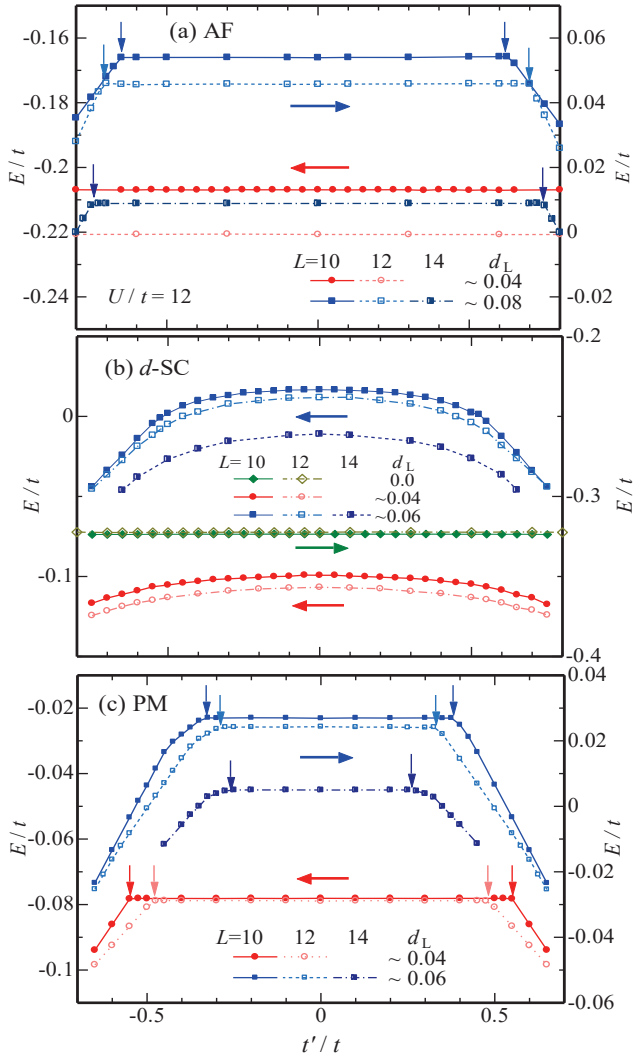} 
%\hskip -4cm
%\vskip -15mm
\end{center} 
\vskip -5mm
\caption{(Color online) 
Total energy as function of $t'/t$ for some $d_{\rm L}$ and $L$ ($U/t=12$) 
in (a) AF, (b) $d$-SC, and (c) PM states. 
For $d_{\rm L}=0$ in (b), we cite previous data with BR2 
(greenish symbols).\cite{SY}
In (a) and (c), the boundaries $\tilde t'/t$ are indicated by arrows 
with the color corresponding to each $L$. 
In every state for $d_{\rm L}>0$, $E/t$ becomes subtly asymmetric with 
respect to $t'/t=0$. 
}
\label{fig:Etot-alpha} 
%\vskip -5mm
\end{figure}
In previous papers,\cite{Watanabe,SY} we showed that the behavior of the 
lowest-energy states ($d_{\rm L}=0$) for $U>U_{\rm c}$ at half filling 
becomes independent of $t'/t$ at least for $|t'/t|\lesssim 0.5$ in various 
states ($d$-SC, AF, staggered-flux, and PM), if BRE is applied [see 
Fig.~\ref{fig:Etot-alpha}(b) below for a $d$-SC case].
In this Appendix, we summarize the effects of the diagonal 
(next-nearest-neighbor) hopping term on the excited states ($d_{\rm L}>0$), 
and argue that $t'/t$ does not affect the essence of this study for 
$|t'/t|\lesssim 0.5$. 
\par
\begin{figure}[t]
\begin{center} 
%\vskip -5mm
\hskip -7mm
\includegraphics[width=8.5cm,clip]{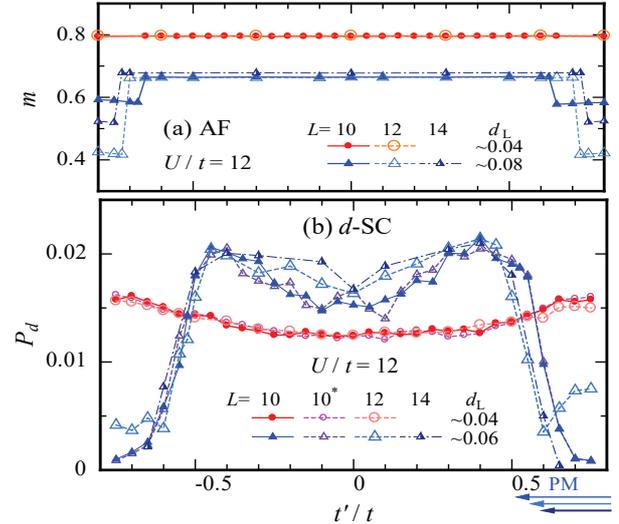} 
%\hskip -0.4cm
%\vskip -15mm
%\includegraphics[width=9.5cm,clip]{d-DF-JPSJ.eps} 
\end{center} 
\vskip -5mm
\caption{(Color online) 
Order parameters of (a) AF state $m$ and (b) $d$-SC state $P_d$ as
functions of $t'/t$ for some values of $d_{\rm L}$ and $L$. 
In (b), data for the two best trials are shown for $L=10$ (as $10$ and 
$10^*$) owing to large statistical errors. 
Judging from the vanishing of $\Delta_d$, the $d$-SC state for 
$d_{\rm L}\sim 0.06$ (bluish symbols) is reduced to the PM state for 
$t'/t\gtrsim 0.5$ as indicated by arrows of bluish colors (corresponding to 
different $L$) with ``PM". 
In this area, $P_d$ should vanish as $L\rightarrow\infty$.\cite{YOTKT}
On the other hand, weak $d$-SC order seems to remain for 
$t'/t\lesssim -0.5$. 
}
\label{fig:Order-para-alpha} 
\vskip -5mm
%1
\end{figure}

Figure \ref{fig:Etot-alpha} shows $t'/t$ dependence of the total energy
in the three states [(a) AF, (b) $d$-SC, and (c) PM] for a few values of 
$d_{\rm L}$ and $L$ and $U/t=12$. 
\par

(1) {\it AF state}: Like the $d_{\rm L}=0$ case, $E/t$ (and the state) 
is unchanging at least for $|t'/t|\le 0.8$ for small $d_{\rm L}$ 
($\lesssim 0.04$). 
Even for relatively large $d_{\rm L}$ ($\sim 0.08$), the optimized 
$\Psi_{d_{\rm L}}^{\rm AF}$ is unchanging for a wide range of $t'/t$ 
($|t'|<|\tilde{t'}|$, $\tilde{t'}$ being the boundary value), and 
$|\tilde{t'}|/t$ increases as $L$ increases, as indicated by arrows in 
Fig.~\ref{fig:Etot-alpha}(a). 
In Fig.~\ref{fig:Order-para-alpha}(a), the staggered magnetization 
is shown; $m$ remains finite even for 
$|t'|>|\tilde t'|$.\cite{note-metallic-AF} 
Thus, the AF state is robust against $t'/t$. 
\par

(2) {\it d-SC state}: $E/t$ slowly changes as $|t'/t|$ increases
for $d_{\rm L}>0$, in contrast to the AF state. 
For large $d_{\rm L}$ ($\gtrsim 0.06$), $E/t$ comes to change rapidly 
for $|t'|/t\gtrsim 0.5$. 
In Fig.~\ref{fig:Order-para-alpha}(b), evolution of the $d$-SC correlation 
function $P_d$ is shown. 
For $|t'|/t\lesssim 0.5$, $P_d$ preserves large values regardless of 
$d_{\rm L}$, whereas for $|t'|/t\gtrsim 0.5$, $P_d$ rapidly drops 
for relatively large $d_{\rm L}$. 
Especially for $t'/t\gtrsim 0.5$ (indicated by arrows), $d$-SC order 
vanishes and the state is reduced to the normal (PM) 
state.\cite{note-dSC}
\par

(3) {\it PM state}: Like the AF state, $E/t$ (and the state) is 
unchanging for $|t'|<|\tilde t'|$ at least for finite $L$. 
However, $|\tilde t'|/t$ decreases as $L$ increases. 
Therefore, for $L\rightarrow\infty$ (as the ${\bf k}$-point mesh 
becomes finer), $E/t$ probably starts decreasing slowly as soon as 
$t'/t$ is introduced like that of $d$-SC. 
Analyzing $n({\bf k})$ (not shown), we found that $t'/t$ dependence of 
$\Psi_{\rm PM}$ is almost limited to the loci of the Fermi surface. 
For $|t'|>|\tilde t'|$, the Fermi surface rapidly deviates from 
the antinodal points, and then $S({\bf q})$ at ${\bf q}={\bf Q}$ 
markedly decreases (not shown). 
This decrease of $S({\bf Q})$ causes the rapid decrease of $P_d$ for large 
$|t'|/t$ shown in Fig.~\ref{fig:Order-para-alpha}(b). 
\par 

\begin{figure}[t]
\begin{center} 
%\vskip -5mm
%\hskip -20mm
\includegraphics[width=8.5cm,clip]{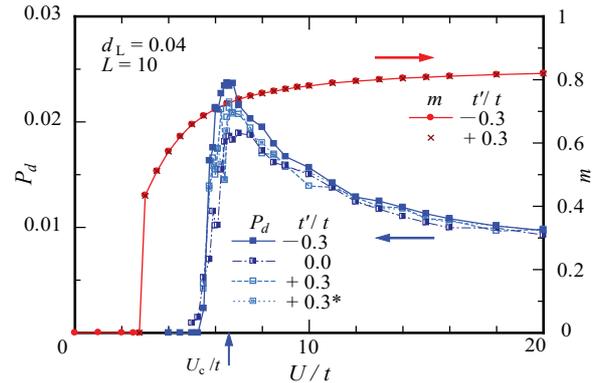} 
\end{center} 
\vskip -5mm
\caption{(Color online) 
$U/t$ dependence of the order parameters $m$ (right axis) and $P_d$ 
(left axis) for $d_{\rm L}=0.04$ are respectively compared among a few 
values of $t'/t=\pm 0.3$. 
For $P_d$ with $t'/t=+0.3$, which has relatively large errors, data of 
the two best trials of different initial conditions (indicated by $+0.3$ 
and $+0.3^*$) are shown for each value of $U/t$. 
The Mott transition point of $\Psi_d$ for $d_{\rm L}=0$ 
($U_{\rm c}/t\sim 6.6$) is indicated by an arrow on the horizontal axis.
}
\label{fig:sp-app} 
\vskip -5mm
%1
\end{figure}

Anyway, the order of energy ($E_{\rm PM}\ge E_{d{\rm-SC}}>E_{\rm AF}$) 
is unchanging by introducing $|t'|/t$ ($\lesssim 0.8$). 
Finally, we mention $U/t$ dependence. 
In Fig.~\ref{fig:sp-app}, the evolution, as $U/t$ is varied, of the order 
parameters $m$ and $P_d$ for excited states ($d_{\rm L}=0.04$) is compared 
among a few values of $t'/t$. 
There is no detectable difference in $m$ ($\lesssim 10^{-3}$) for any $U/t$, 
and $P_d$ exhibits only a slight quantitative difference. 
\par

To summarize, $t'/t$ dependence in the excited states is only quantitative 
for $|t'|/t\lesssim 0.5$; 
The properties for $t'/t=-0.3$ discussed in the main text are essentially 
unchanging for other moderate values of $t'/t$. 
\par

%============================================================================


\begin{thebibliography}{99}
\bibitem{mixed} Although it is possible to use a mixed state of AF and SC 
orders, we leave such an integrated state for future studies.
\bibitem{note-min} 
It is not easy to perfectly remove the possibility of overlooking 
the true global minimum, although we carefully monitored every VMC sweep.
\bibitem{note-YOT} In Ref.~\citen{YOT}, in which BRE was not considered, 
the Mott transition point $U_{\rm c}/t$ shifts to large values as 
$|t'/t|$ ($t'/t<0$) increases. 
If BRE is properly introduced, $U_{\rm c}/t$ becomes independent of $|t'/t|$ 
($\lesssim 0.5$): $U_{\rm c}/t\sim 8.5$.
\bibitem{note-N(q)}
Within the single-mode approximation, $N({\bf q})$ behaves for 
$|{\bf q}|\rightarrow 0$ as $\propto|{\bf q}|$ if the state is gapless 
in the charge sector, whereas 
$N({\bf q})\propto|{\bf q}|^\alpha~~(\alpha\ge 2)$ if a gap 
opens.\cite{Auerbach}
\bibitem{note-binding} This tendency is consistent with that of the 
previous studies\cite{Takahashi1,Takahashi2} for the $t$-$J$-type model, 
arguing that the D-H correlation between the nearest-neighbor sites is 
repulsive. 
In the present case, however, the D-H correlation remains attractive 
($\zeta>0$), because it is affected by the part of D and H already existing 
in the lowest-energy state ($d_{\rm L}=0$), which is strongly attractive 
as shown in Fig.~\ref{fig:para-zeta} (red symbols).
\bibitem{note-dF} By considering the system-size dependence, $d_{\rm F}$ 
for $U>U_{\rm c}$, which is finite for finite $L$, completely vanishes 
in the limit of $L\rightarrow\infty$.
\bibitem{note-FD} The reason why free doublons appear and increase 
as $d_{\rm L}$ increases is considered as follows. 
As $d$ approaches $d_{\rm L}$ (for large $U/t$), it becomes difficult 
that the total energy is reduced by creating and, especially, destroying 
D-H pairs (namely, by $E_{\rm kin}^{({\rm L})}$), to keep the regulation 
$d\ge d_{\rm L}$. 
To reduce the total energy, it becomes advantageous that D-H pairs are 
resolved and hopping of free doublons and holons arising thereby reduces 
$E_{\rm kin}^{({\rm F})}$.
\bibitem{note-Pd} It is known\cite{YOTKT} that $P_d$ measured using 
$\Psi_0^{({\rm PM})}$ (or $\Psi_0^{d{\rm -SC}}$ with $\Delta_d=0$) becomes 
finite for finite $L$ but converges to zero for $L\rightarrow\infty$; 
actually, $P_d$ looks almost vanishing for $L$ as large as $30$ (see 
Fig.~C$\cdot$1 in Ref.~\citen{YOTKT}). 
Therefore, to avoid misunderstanding, we plot $P_d$ as zero on purpose 
in the case where the optimized $\Delta_d$ becomes substantially zero, 
namely, $\Psi_d$ is incoherent, in this article. 
Do not confuse it with a case of $P_d=0$ with large $\Delta_d$, as 
on the Mott-insulator side (small $d_{\rm L}$ for $U\gtrsim U_{\rm c}$).
\bibitem{note:d-dL} Equation (\ref{eq:d-dL}) is approximately valid 
in the $d$-SC case, but $d_{\rm Lc}$ is somewhat dependent on $U/t$ 
and $L$, in contrast with $d_{\rm c}$. 
On the other hand, $d_{\rm LM}$ seems less dependent on $U/t$ than 
$d_{\rm M}$.
\bibitem{note-metallic-AF} 
For $|t'|>|\tilde t'|$, $\Psi_{\rm AF}$ becomes metallic with finite $m$. 
This state is similar to the carrier-doped AF state,\cite{SY} in which 
hole-pocket-type Fermi surfaces arise near $(\pi,0)$ for $t'/t>0$ and 
near $(\pi/2,\pi/2)$ for $t'/t<0$. 
In the present case, because plus-charged (holons) and minus-charged 
(doublons) carriers are simultaneously generated, hole-pocket-type 
Fermi surfaces arise near $(\pi,0)$ for $t'/t>0$ [$(\pi/2,\pi/2)$ 
for $t'/t<0$] and electron-pocket-type Fermi surfaces simultaneously 
arise near $(\pi/2,\pi/2)$ for $t'/t>0$ [$(\pi,0)$ for $t'/t<0$].
\bibitem{note-dSC} In this area ($t'/t\gtrsim 0.5$, 
$d_{\rm L}\gtrsim 0.06$), we confirmed that $\Delta_d$ vanishes 
(not shown). 
For $t'/t\lesssim -0.5$, weak $d$-SC seems to remain ($\Delta_d>0$).
\end{thebibliography}
\end{document}